\documentstyle[12pt]{article}

\newcommand{\Frac}[2]%
{{\textstyle \frac{\mbox{\footnotesize $#1$}\rule[-0.9mm]{0mm}{1mm}}%
{\mbox{\footnotesize $#2$}\rule{0mm}{3.1mm}}}}

\def\cxx{1 - {{4x^{2} P^{2}}\over{Q^{2}}} }
\def\cx{1 - {{2x    P^{2}}\over{Q^{2}}} }

\def\cut{\sqrt{1 - {{4(m^{2}+\lambda^2)}\over {s}} } }

\begin{document}
\begin{flushright}
SLAC--PUB--7819\\
MPI H-V17-1998\\
May 1998
\end{flushright}
\bigskip
\vfill

\begin{center}
{\Large \bf The Spin Structure of a Polarized Photon}
  \footnote{\baselineskip=14pt
     Work supported in part by the Department of Energy, contract 
     DE--AC03--76SF00515 and by Fondecyt (Chile) under grant 1960536 and by a
C\'atedra Presidencial (Chile).}
\vspace{22pt}

  \centerline{\bf Steven D. Bass}
\vspace{8pt}
  \centerline{\it Max Planck Institut f\"ur Kernphysik}
  \centerline{\it Postfach 103980, D-69029 Heidelberg, Germany}
  \centerline{e-mail: Steven.Bass@mpi-hd.mpg.de}
\vspace*{0.9cm}

  \centerline{\bf Stanley J. Brodsky}
\vspace{8pt}
  \centerline{\it Stanford Linear Accelerator Center}
  \centerline{\it Stanford University, Stanford, California 94309}
  \centerline{e-mail: sjbth@slac.stanford.edu}
\vspace*{0.9cm}

  \centerline{\bf Ivan Schmidt}
\vspace{8pt}
  \centerline{\it Departmento de F\i sica, Universidad T\'ecnica Federico Santa Mar\i a}
  \centerline{\it Casilla 110-V, Valpara\i so, Chile}
  \centerline{e-mail: ischmidt@newton.fis.utfsm.cl}
\vspace*{0.9cm}

\end{center}

\begin{abstract}
\noindent
We show that the first moment of the spin-dependent structure function
$g_1^{\gamma}(x,Q^2)$
of a real photon vanishes independent of the
momentum transfer $Q^2$ it is probed with.
This result is non-perturbative:
it holds to all orders in perturbation theory in
abelian and non-abelian gauge theory and at
every twist.
\end{abstract}
\vfill
\newpage

\section {A sum-rule for $g_1^{\gamma}$}

Polarized photon-photon collisions offer a new laboratory for studying
QCD spin physics.
In polarized deep inelastic scattering the spin-dependent
structure function $g_1^{\gamma}(x,Q^2)$ of a polarized photon
\cite{ahmed,manohar,ijmp}
is sensitive to the axial anomaly \cite{ijmp,efremov}
and thus to the realization of chiral symmetry in QCD
\cite{shore,sehgal}.
The spin-dependent parton distributions of the polarized photon could
be measured
in photoproduction studies with a polarized proton beam at HERA
\cite{strat,hera}.
Polarized real photon collisions could be studied
with high-energy real photon beams at the NLC \cite{twop,nlcg}.

A remarkable feature of polarized deep inelastic
scattering for ($Q^2 \rightarrow \infty$) on a real photon target
is that the leading twist (=2) contribution
to the first moment of $g_1^{\gamma}$
vanishes\cite{ijmp}.
This (deep inelastic) result is nonperturbative and follows
directly from electromagnetic gauge invariance and the absence of
any exactly massless Goldstone boson in the physical spectrum.
In addition, it has  recently been shown\cite{brod97} that the first
moment of the box graph contribution to polarized $\gamma \gamma$
fusion vanishes when one or both of the incident
photons is real -- independent of the virtuality of the second photon.
In this paper we generalize these two results and show that the
first moment of $g_1^{\gamma}$ for a real photon vanishes to all
orders and at every twist.

Consider polarized $\gamma - \gamma$ scattering where $\sigma_A$
and $\sigma_P$ denote the two cross-sections for the absorption
of a transversely polarized photon with spin anti-parallel
$\sigma_A$ and parallel $\sigma_P$ to the spin of the target
photon.
The photons in a lepton-lepton collider can be real or spacelike.
We let $q_{\mu}$ and $p_{\mu}$ denote the
momentum of the  ``incident'' and ``target'' photons
and define $Q^2 = -q^2$, $P^2 = -p^2$ and $\nu = p.q$.
The spin-dependent part of the total $\gamma \gamma$ cross-section
is given by
\begin{equation}
(\sigma_A - \sigma_P) = {8 \pi^2 \alpha \over {\cal F}} g_1^{\gamma}
(Q^2, \nu, P^2) .
\end{equation}
Here $g_1^{\gamma}$ is the target photon's spin-dependent structure
function,
and ${\cal F}$ is the flux factor for the incident photon.
The flux factor is discussed in Eqs. (5-8) below.
The structure function $g_1^{\gamma}(Q^2,\nu,P^2)$
is symmetric under the exchange of the incident and target photons
$(p \leftrightarrow q)$.
There is no $g_2$ contribution to $(\sigma_A - \sigma_P)$
\cite{ahmed,manohar}.

Now consider a real photon beam: $Q^2=0$.
The Drell-Hearn-Gerasimov sum-rule \cite{dhg}
(--- for a review see \cite{moia})
for spin-dependent photoproduction
tells us that
the integral
$\int_0^{\infty} {d \nu \over \nu} (\sigma_A - \sigma_P)$
is proportional to $\alpha$ times the square
of the (photon) target's anomalous magnetic moment.
The Drell-Hearn-Gerasimov sum-rule is derived
from the dispersion relation for the spin-dependent part
of the forward Compton amplitude.\footnote{
The Drell-Hearn-Gerasimov sum-rule is derived for QED and QCD with
a finite mass gap (massive fermions).
The dispersion relation for the spin-dependent part $f_2(\nu)$ of
the forward Compton amplitude relates the integral on the right
hand side of Eq. (2) to the first derivative of the real part of
$f_2(\nu)$ evaluated at $\nu \rightarrow 0$.
Provided that there is a finite mass gap between the ground state
and continuum contributions to forward Compton scattering,
when we take the low energy limit that $\nu \rightarrow 0$
the leading term in Re$f_2(\nu)$
is proportional to $\nu$
times the square of the target's anomalous magnetic moment
\cite{low,brod69}.}
It follows from the general principles of causality,
unitarity, Lorentz and electromagnetic gauge invariance
and the assumption
that $g_1^{\gamma}$ satisfies an unsubtracted dispersion
relation.
Modulo this no-subtraction hypothesis,
the Drell-Hearn-Gerasimov sum-rule is valid for a target of arbitrary
spin $S$,
whether elementary or composite \cite{brod69}.

For a real incident photon the flux factor ${\cal F} = \nu$.
Furry's theorem tells us that the photon has zero anomalous magnetic
moment (both in QED and in QED coupled to QCD).
It follows that
\begin{equation}
\int_0^{\infty} {d \nu \over \nu} (\sigma_A - \sigma_P)
=
8 \pi^2 \alpha
\int_{\nu_{th}}^{\infty} {d \nu \over \nu} {g_1^{\gamma} \over \nu}
= 0,
\ \ \ \ \ \ (P^2 = Q^2 = 0) .
\end{equation}
Here $\nu_{th}$ is the threshold energy: $\nu_{th} = 2m_e^2$ in QED
and $\nu_{th} = {1 \over 2} m_{\pi}^2$ in QCD.
Eq. (2) is a non-perturbative result.
It holds to all orders in perturbation theory in both QED and QCD.
If we replace the photon target by a $W^{\pm}$ boson target,
then
the Drell-Hearn-Gerasimov integral (2) is finite starting at
${\cal O}(\alpha^3)$ since the $W^{\pm}$ boson has a finite
anomalous magnetic
moment starting at ${\cal O}(\alpha)$ \cite{brodsc}.

We now generalize this result to the case where one of
the two photons becomes virtual: $Q^2 > 0$.
Furry's theorem implies that the anomalous magnetic moment of
a photon vanishes independently of whether the photon is real
or virtual.
Since $g_1^{\gamma}$ and $\nu$ are each symmetric under
the exchange of ($p \leftrightarrow q$),
we can treat the virtual photon as the target and the real
photon as the beam,
and then apply the Drell-Hearn-Gerasimov sum-rule to find
\begin{equation}
I^{\gamma}(Q^2) \equiv
\int_{\nu_{th}}^{\infty} {d \nu \over \nu} {g_1^{\gamma}(\nu, Q^2,P^2) \over
\nu} = 0
\end{equation}
independent of $Q^2$ provided that $P^2 =0$.
Changing the integration variable from $\nu$
to Bjorken $x={Q^2 \over 2 \nu}$,
we can rewrite Eq. (3) as
\begin{equation}
I^{\gamma}(Q^2) = {2 \over Q^2} \int_0^{x_{max}} dx g_1^{\gamma} (x,Q^2,P^2=0)
=0
\ \ \ \ \ \ \forall Q^2 .
\end{equation}
The threshold factors in Eqs. (3) and (4) are
$\nu_{th}=(Q^2+4m_e^2)/2$ and $x_{max} = Q^2/(Q^2+4m_e^2)$
in QED,
and
$\nu_{th}=(Q^2+m_{\pi}^2)/2$ and $x_{max} = Q^2/(Q^2+m_{\pi}^2)$
in QCD.

The function $I^{\gamma}(Q^2)$ interpolates between $Q^2=0$ and
polarized deep inelastic scattering.
The corresponding integral for a nucleon target was introduced
previously
by Anselmino, Ioffe and Leader in \cite{ansel}.

Equations (3) and (4) give our main result.
A corollary is that $g_1^{\gamma}$
must change sign at least once at a value $x=x^*(Q^2)$ since the first moment of
$g_1^{\gamma}$ vanishes. The crossing point $x^*$ for the box graph
contribution  to polarized
$\gamma \gamma$ fusion has been calculated in \cite{brod97}.

It is important to note that the new sum-rule (4) involves
$g_1^{\gamma}$ instead of $(\sigma_A - \sigma_P)$.
For real incident photons the flux factor ${\cal F}$ is equal
to $\nu = p.q$.
For virtual incident photons the flux factor is
convention dependent
subject to the requirement that
\begin{equation}
\lim_{Q^2 \rightarrow 0} {\cal F} = \nu .
\end{equation}
There are two popular choices due to Gilman \cite{gilman} and
Hand \cite{hand} which are used in virtual-photon nucleon
collisions.   Both of these conventions readily generalize to
photon targets as follows:
\begin{equation}
{\cal F}_{\rm Gilman} = \sqrt{\nu^2 + P^2 Q^2}
\end{equation}
and
\begin{equation}
{\cal F}_{\rm Hand} = \nu (1 - x) .
\end{equation}
In a recent paper \cite{brod97}, Brodsky and Schmidt
have employed:
\begin{equation}
{\cal F}_{\rm BS} = \nu = {1 \over 2} (s + Q^2 + P^2) .
\end{equation}
The Gilman and the Brodsky-Schmidt conventions preserve
the $(p \leftrightarrow q)$ symmetry between the target and
incident photons
whereas the generalized Hand convention does not.
Using ${\cal F}_{\rm BS}$, Brodsky and Schmidt \cite{brod97}
discovered that the box graph, ${\cal O}(\alpha^2)$,
contribution to $(\sigma_A - \sigma_P)$ in polarized
photon-photon fusion satisfies Eq. (2) with $Q^2 >0$.
The sum-rule (4) generalizes their result to all orders.

In the remainder of this paper we explore the symmetry properties of the
box graph contribution to $g_1^{\gamma}$ after we impose various
kinematic cut-offs to separate the total phase space into ``hard''
and ``soft'' contributions.
We discuss the application of these symmetry arguments to factorization
in the QCD parton model.
We then use the ($p \leftrightarrow q$) symmetry of $g_1^{\gamma}$ to show that
Eq. (4) holds twist by twist in polarized deep inelastic scattering.
Finally, we extend our results to the {\it gedanken} world of massless
quarks in QCD where the Drell-Hearn-Gerasimov sum-rule is not guaranteed to
hold.

\section{($p \leftrightarrow q$) symmetry and photon-photon fusion}

Consider the box graph contribution to photon-photon fusion.
It is illuminating to evaluate the box graph with a cut-off on the transverse
momentum squared of the struck quark relative to the photon-photon direction:
$k_T^2
\geq
\lambda^2$. The cut-off separates the total phase space into ``hard''
($k_T^2 \geq \lambda^2$) and ``soft''
($k_T^2 < \lambda^2$) contributions.
One finds \cite{bnt}:
\begin{eqnarray}
g_1^{\gamma} (x,Q^{2},P^{2})|_{\rm hard} &=&
-{\alpha \over \pi }
{\cut \over \cxx} \Biggl[ (2x-1)(\cx) \\ \nonumber
& &
\biggl(1 - {1 \over {\cut \sqrt{\cxx} }}
\ln \biggl({ {1+\sqrt{\cxx} \cut}\over {1-\sqrt{\cxx} \cut}}
\biggr) \biggr) \\ \nonumber
& &
+ (x-1+{{x P^{2}}\over{Q^{2}}})
{{\left( 2m^{2}(\cxx)- P^{2}x(2x-1)(\cx)\right)}
\over {(m^{2} + \lambda^2) (\cxx) - P^{2}x(x-1+{{x P^{2}}\over{Q^{2}}})}}
\Biggr]
\end{eqnarray}
for each type of fermion liberated into the final state
\footnote{
Quark contributions to $g_1^{\gamma}$ are obtained by multiplying
the right hand side of Eq. (9) by the number of colors ($N_c=3$).}
.
Here
$m$ is the fermion mass,
$x$ is the Bjorken variable ($x= {Q^2 \over 2 \nu}$)
and
$s$ is the center of mass energy squared
\begin{equation}
s= (p+q)^2 = Q^2 \biggl( {1 - x \over x} \biggr) - P^2
\end{equation}
for the photon-photon collision.
In perturbative QCD the box graph contribution to the spin structure
function of a polarized gluon
$g_1^{(g)}(x,Q^2,P^2)$
for $k_T^2 \geq
\lambda^2$ is  obtained from Eq. (9) by substituting
${\alpha \over \pi}$
by ${\alpha_s \over 2 \pi}$.

In general, the cut-off $\lambda^2$ may be chosen to be a function of
$x$ [16-20]:
\begin{equation}
\lambda^2 = \lambda_0^2 f_0(x) + P^2 f_1(x) + m^2 f_2(x) .
\end{equation}
If we set $\lambda^2$ to zero, thus including the entire phase space,
then we obtain the full box graph contribution to $g_1^{\gamma}$.
If we take $\lambda^2$ to be finite and independent of $x$,
then the crossing symmetry of
$g_1^{\gamma}$ under the exchange of ($p \leftrightarrow q$)
is realized separately in each of the ``hard'' and ``soft''
parts of $g_1^{\gamma}$ which correspond to phase space with
($k_T^2 > \lambda^2$) and ($k_T^2 < \lambda^2$) respectively.
We could also choose an $x$-dependent cut-off on the struck
quark's virtuality
\cite{bnt,bint}
\begin{equation}
m^2-k^2 = P^2 x + {k_T^2+m^2\over (1-x)} > \lambda_0^2 = {\rm constant}(x)
\end{equation}
or a cut-off on the invariant mass squared
of the quark-antiquark
component of
the light-cone wavefunction of the target photon \cite{mank,lepage}
\begin{equation}
{\cal M}^2_{q {\overline q}} = {k_T^2 + m^2 \over x (1-x)} + P^2
\geq \lambda_0^2 = {\rm constant}(x) .
\end{equation}
Substituting Eqs. (11-13) into Eq. (9) we find that the ``hard''
and ``soft'' contributions to $g_1^{\gamma}$ do not separately
satisfy the
$(p \leftrightarrow q)$ symmetry of $g_1 (x,Q^{2})$ if use an
$x$-dependent cut-off to define the ``hard'' part of the total
phase space.
The reason for this is that the transverse momentum is defined
perpendicular to the plane spanned by $p_{\mu}$ and $q_{\mu}$
in momentum space.
The $x$-dependent cut-offs mix the transverse and longitudinal
components of momentum.
They induce a violation of crossing symmetry
in
$g_1|_{\rm hard} (x,Q^{2},P^{2})$ under $(p \leftrightarrow q)$.

If we set $P^2$ and $\lambda^2$ to zero in Eq. (9) we obtain
the box graph contribution to $g_1^{\gamma}$ for a real
photon target:
\begin{equation}
g_1^{\gamma} =
   - {\alpha \over \pi}     \sqrt{1 - {4m^2 \over s}}
   \biggl[ (2x-1) \biggl( 1 -   { 1 \over \sqrt{1 - {4m^2 \over s}} }
        \ln \biggl( { { 1 + \sqrt{1 - {4m^2 \over s}} }
            \over   { { 1 - \sqrt{1 - {4m^2 \over s}} } } } \biggr)
        +2 (x-1)
        \biggr) \biggr] .
\end{equation}
The discovery in Ref.\cite{brod97} is that Eq. (4) vanishes for the box
graph contribution --- Eq. (14).
The structure function $g_1^{\gamma}$ in Eq. (14) can be written as the
sum of two contributions $g_1^{\gamma}|_{\rm like}$ and
$g_1^{\gamma}|_{\rm unlike}$ where the two fermions in the final state
have the same spin ($g_1^{\gamma}|_{\rm like}$)
and opposite spins ($g_1^{\gamma}|_{\rm unlike}$).
Working in the limit $Q^2 \gg m^2$,
Freund and Sehgal \cite{sehgal} have found that the first moments of
$g_1^{\gamma}|_{\rm like}$
and
$g_1^{\gamma}|_{\rm unlike}$ yield the explicit and anomalous chiral
symmetry breaking contributions to the photon's axial charge.
These two contributions cancel in the deep inelastic limit
($P^2 \ll m^2 \ll Q^2$) \cite{ccm,bint}.

The $(p \leftrightarrow q)$ symmetry of $g_1|_{\rm hard}(x,Q^{2},P^{2})$
has application to the QCD parton model.
The parton model description of polarized deep inelastic scattering
involves writing the deep inelastic structure functions as the sum
over the convolution of ``soft'' quark and gluon parton distributions
with ``hard'' photon-parton scattering coefficients.
The flavor-singlet part of $g_1$ may be written
\begin{equation}
g_1|_{\rm singlet} = {1 \over 9}
\biggl( \sum_q \Delta q \otimes C^q + N_f \Delta g \otimes C^g
\biggr)
.
\end{equation}
Here, $\Delta q$ and $\Delta g$ denote the quark and gluon parton
distributions, $C^q$ and $C^g$ denote the corresponding hard
scattering coefficients, and $N_f$ is the number of quark flavors
liberated into the final state.
The parton distributions are target dependent and describe a flux of
quark and gluon partons into the hard (target independent)
photon-parton interaction which is described by the coefficients.
The separation of $g_1$ into ``hard'' and ``soft'' is not unique
and
depends on the choice of factorization scheme [16-20].

We can use the kinematic cut-off on the
partons' transverse momentum  squared $k_T^2$ to define the
factorization scheme and
thus separate the hard and soft parts of the phase  space for
the photon-parton collision. Following Eq. (11), this cut-off
may be $x$-dependent or
$x$-independent. In the QCD parton model $g_1^{(g)}|_{\rm hard}(x,Q^2)$
is a suitable candidate for the hard coefficient $C^g$
in photon-gluon fusion.
Among the possible kinematic cut-offs,
the $x$ independent cut-off
on the
transverse momentum squared
preserves the crossing symmetry of $g_1^{(g)}$ under
$(p \leftrightarrow q)$
in both the hard gluonic coefficient
$C^g = g_1^{(g)}|_{\rm hard}(x,Q^2)$ and the soft polarized quark
distribution of the gluon
$\Delta q^{(g)} = g_1^{(g)}|_{\rm soft}(x,Q^2)$.

The $x$-independent cut-off is especially suited to discussions
about the axial anomaly in polarized deep inelastic scattering.
Suppose that
we evaluate the box graph contribution to the first moment
of
$g_1^{(g)}$ with an $x$-independent cut-off:
$k_T^2 > \lambda^2$ where ($m^2, P^2 \ll \lambda^2 \ll Q^2$).
Then, we find the axial-anomaly \cite{adler,bell}
as a contact photon-gluon interaction associated
with
$k_T^2 \sim Q^2$ \cite{ccm}.
On the other hand,  the first moment of
$g_1^{(g)}$, defined using the quark virtuality $(-k^2)$ cut-off
 yields ``half of
the anomaly''  in the gluon coefficient
through the mixing of transverse and longitudinal momentum
components \cite{bnt,bint}.
The anomaly coefficient for the first moment is recovered with the
invariant mass squared cut-off through a sensitive cancelation of
large and
small $x$ contributions \cite{bint}.

\section{Twist expansion for $g_1^{\gamma}$ when $Q^2 \rightarrow \infty$}

The light-cone operator product expansion  at  large $Q^2$
relates the first moment of the structure function $g_1^{\gamma}$ to the
scale-invariant axial charges of the target photon
\cite{ahmed,manohar,ijmp}
plus an expansion of higher-twist matrix elements:
\begin{eqnarray}
& & \int_0^1 dx  g_1^{\gamma}(x,Q^2,P^2) \\ \nonumber
& & = \Biggl( {1 \over 12} a^{(3)} + {1 \over 36} a^{(8)} \Biggr)
    \Bigl\{1 + \sum_{\ell\geq 1} c_{{\rm NS} \ell\,}
    \bar{g}^{2\ell}(Q)\Bigr\} + {1 \over 9} a^{(0)}|_{\rm inv}
    \Bigl\{1 + \sum_{\ell\geq 1} c_{{\rm S} \ell\,}
    \bar{g}^{2\ell}(Q)\Bigr\}
\\ \nonumber
& & \ \ +
\sum_{j=1}^{\infty} \biggl( {P^2 \over Q^2} \biggr)^j
          \{ {\rm twist} \ (2 + 2j) \ {\rm operator \ matrix \ elements} \}
%           \{\rm x \ coefficients}(Q^2)
\\ \nonumber
& & \ \ +
\sum_{j=1}^{\infty} \biggl( {m^2 \over Q^2} \biggr)^j \
\sum_{k=0}^{\infty} \ \biggl( {P^2 \over Q^2} \biggr)^k
          \{ {\rm twist} \ (2 + 2k) \ {\rm operator \ matrix \ elements} \}
.
\end{eqnarray}
where $m$ is the quark mass.
(We refer to \cite{maul} for a complete derivation of the twist-4
contributions to deep inelastic scattering from a nucleon target.)

For photon states $|\gamma(p,\lambda) \rangle$ with momentum $p_\mu$
and polarization $\lambda$
\begin{equation}
i a^{(k)} \epsilon_{\mu \nu \alpha \beta} p^{\nu} \epsilon^{\alpha}(\lambda)
 \epsilon^{* \beta}(\lambda)
=
\langle \gamma (p,\lambda) |
J_{\mu 5}^{(k)}
| \gamma (p,\lambda) \rangle _c  \nonumber \\
\end{equation}
where $k=(3,8,0)$ and the subscript $c$ denotes the connected matrix element.
The non-singlet isovector and SU(3) octet currents are
\begin{eqnarray}
J_{\mu 5}^{(3)} &=&
\left(\bar{u}\gamma_\mu\gamma_5u - \bar{d}\gamma_\mu\gamma_5d \right)
\nonumber \\
J_{\mu 5}^{(8)} &=&
\left(\bar{u}\gamma_\mu\gamma_5u + \bar{d}\gamma_\mu\gamma_5d
                   - 2 \bar{s}\gamma_\mu\gamma_5s\right)
\end{eqnarray}
and
\begin{equation}
J^{(0)}_{\mu5} =
%&=&
E(g) \
\left(\bar{u}\gamma_\mu\gamma_5u
                  + \bar{d}\gamma_\mu\gamma_5d
                  + \bar{s}\gamma_\mu\gamma_5s\right)_{GI}
\end{equation}
is the
scale invariant and gauge-invariantly renormalized singlet axial-vector
operator.
The renormalization group factor $E(g)$ \cite{mink}
compensates for the non-zero anomalous dimension \cite{koeb,rjc,collins,kod}
of the singlet axial-vector current
$J^{(0)}_{\mu5}/E(g)$.
The flavor non-singlet $c_{{\rm NS} \ell}$
and singlet $c_{{\rm S} \ell}$ coefficients
are calculable in
$\ell$-loop perturbation theory \cite{larin}.
There are no twist-two, spin-one, gauge invariant photon or gluon operators
which can contribute to the first moment of $g_1^{\gamma}$ \cite{jaffe}.

One can derive a rigorous sum-rule for the leading twist (=2) contribution
to the first moment of $g_1^{\gamma}$ in polarized deep inelastic
scattering where one of the photons is deeply virtual
($Q^2 \rightarrow \infty$)
and the other photon is either real \cite{ijmp} or carries small
but finite virtuality \cite{shore}.
Electromagnetic gauge-invariance implies \cite{ijmp}
that the axial charges of a real photon vanish provided that
there is no exactly massless Goldstone boson coupled to
$J_{\mu 5}^{(k)}$, which is certainly true in nature with massive quarks.
For real photons we find \cite{ijmp}:
\begin{equation}
\int_0^1 dx g_1^{\gamma}|_{\{\rm twist \ 2\}} (x,Q^2,P^2) = 0,
\ \ \ \ \ \ \ \ (P^2=0, \ Q^2 \rightarrow \infty) .
\end{equation}
This deep inelastic sum-rule holds at every order in perturbation
theory -- starting with the box graph for photon - photon fusion.
Comparing Eqs. (20) and (4)
we find that
the vanishing of the leading twist contribution
to
$\int_0^1 dx g_1^{\gamma}$ is a special case of
the Drell-Hearn-Gerasimov sum-rule
when the real photon is treated as the beam and the deeply virtual
photon is treated as the target.

We now consider the higher-twist terms.

The higher twist terms receive contributions from both the ``handbag''
and ``cat-ears'' diagrams.
To classify these terms we note that there are five scales in the
physical problem: $Q^2$, $P^2$, $\nu$, the quark mass $m$ and a QCD
scale $\Lambda$ associated with non-perturbative bound-state dynamics.
We integrate over the scale $\nu$ when we evaluate the first moment
of $g_1^{\gamma}$.

To understand the higher-twist terms in Eq. (16) it is helpful
to first consider the abelian QED contributions to $g_1^{\gamma}$.
There are higher-twist terms proportional to non-zero powers
of ${P^2 \over Q^2}$ and ${m_e^2 \over Q^2}$.
The terms proportional to ${P^2 \over Q^2}$ vanish for a real
photon target ($P^2=0$).
The higher-twist terms proportional to ${m_e^2 \over Q^2}$ start
with the leading twist (=2) operator matrix element.
Fermion mass terms make a non-leading contribution to the Dirac trace
over $\gamma_{\mu}$ matrices when we evaluate $g_1^{\gamma}$ to any
given order in $\alpha$.
They yield a unity matrix contribution to the trace so that
the leading term in the Dirac trace is
the twist-two operator matrix element.
Since the photon's axial charges $a^{(k)}$ vanish when $P^2=0$
it follows that
the higher-twist contributions to (16) vanish for a real photon
target in QED.

In QCD we also have to consider the possible effects of $\Lambda$ and
whether we can have  higher-twist terms proportional to
${\Lambda^2 \over Q^2}$ beyond the higher-twist terms listed in
Eq. (16).
This would also include vector meson dominated contributions to the
cross section.

If we could calculate $g_1^{\gamma}$ exactly in QCD, we would find
an  expression which is symmetric under $(p \leftrightarrow q)$.
This symmetry imposes strong constraints on the possible
$\Lambda$ dependence of $g_1^{\gamma}$.
As an example, recall that the box contribution $g_1^{\gamma}|_{\rm hard}$
in Eq. (9) is symmetric under $(p \leftrightarrow q)$ only with a special
choice of infrared cut-off (independent of $x$).
If we impose the physically sensible condition of not allowing $\Lambda^2$
to scale with the kinematic variables,
then we find that any higher-twist contribution involving $\Lambda^2$ comes
from rescaling the quark mass in one or more terms in the complete QCD
expression for $g_1^{\gamma}$, viz. $m^2 \rightarrow (m^2 + \Lambda^2)$.
That is, if there are higher-twist terms in $g_1^{\gamma}$ proportional
to ${\Lambda^2 \over Q^2}$,
then they effectively induce a constituent-quark mass-term in the higher-twist
expansion. These higher-twist terms thus also vanish for $P^2=0$ because the
photon's axial charges vanish on-shell.

\section{Massless QCD}

It is interesting to extend our results to QCD with massless quarks.
If we could turn the up, down and strange quark masses
to zero in QCD, then the pion and the $\eta$ would evidently become
massless but, because of $U_A(1)$ dynamics \cite{crewther},
the ${\eta'}$ would remain massive.
Consider the gedanken world of massless QCD where we define
real photons by
first taking the light-quark masses to zero and
then
taking the photon virtuality to zero
--- that is, working in the limit $m^2 \ll P^2 \rightarrow 0$.
In this {\it gedanken} world the real-photon's isotriplet
$a^{(3)}$ and octet $a^{(8)}$ axial-charges would
no longer vanish
but instead would be equal to $-{\alpha \over \pi}N_c$
where $N_c=3$ is the number of colors \cite{ijmp,shore}.
The singlet axial-charge $a_0|_{\rm inv}$
would remain zero
since the photon matrix elements of
$J_{\mu 5}^{(0)}$
would not contain a massless pole contribution (because of the massive
$\eta'$). The non-vanishing of the non-singlet $a^{(k)}$ in massless QCD
does not
contradict  our general result (4) because, even for the photon-photon fusion
process (9), the low energy theorem \cite{low,brod69} which relates
the Drell-Hearn-Gerasimov integral to the (vanishing)
anomalous magnetic moment of the target photon
is derived assuming that the fermions have a finite mass.

Gorskii, Ioffe, and Khodjamirian \cite{ioffe}
have found a similar, anomalous, result
in unpolarized photon-photon scattering.
Consider the box graph cross-section
for a hard transverse
photon $\gamma_T$ with virtuality $Q^2$ to scatter from a
soft
longitudinal
photon $\gamma_L$ with virtuality $P^2$:
$\gamma_T \gamma_L \rightarrow l^+ l^-$
where $l$ is the charged fermion liberated into the final state.
This cross section vanishes when we take $P^2 \rightarrow 0$ in
QED with a finite mass gap (the fermion mass $m \neq 0$) and
also in the particular chiral limit
$P^2 \ll m^2 \rightarrow 0$.
However, the $\gamma_T \gamma_L$ cross-section is finite and
non-vanishing in the alternative chiral limit defined by
$m^2 \ll P^2 \rightarrow 0$.

\section{Conclusions}

We have shown that the first moment of the structure
function $g_1^{\gamma}(x,Q^2,P^2)$ measured in  polarized photon-photon
collisions $\gamma(p) \gamma(q) \to X$ vanishes when either or both of the
incident photons are on-shell. This sum rule follows from the
Drell-Hearn-Gerasimov sum rule and simple $ p \leftrightarrow q$  symmetry
properties of the two-photon system. It holds in QED and QCD  to all orders and
at every twist provided that the fermions in the theory have non-vanishing
mass.

\section*{\bf Acknowledgments}

We thank A. H. Mueller for helpful conversations. This work was
supported in part by the United States Department of Energy  under contract
number DE--AC03--76SF00515, by
Fondecyt (Chile) under grant 1960536, and by a C\'atedra Presidencial (Chile).


\begin{thebibliography}{99}
%
\bibitem{ahmed}
M.A. Ahmed and G.G. Ross, Phys. Lett. {\bf B59} (1975) 369.
%
\bibitem{manohar}
A.V. Manohar, Phys. Lett. {\bf B219} (1989) 357.
%
\bibitem{ijmp}
S.D. Bass, Int. J. Mod. Phys. {\bf A7} (1992) 6039.
%
\bibitem{efremov}
A.V. Efremov and O.V. Teryaev, Phys. Lett. {\bf B200} (1990) 200.
%
\bibitem{shore}
S. Narison, G.M. Shore and G. Veneziano, Nucl. Phys. {\bf B391}
(1993) 69;  \\
G.M. Shore and G. Veneziano, Mod. Phys. Lett. {\bf A8} (1993) 373.
%
\bibitem{sehgal}
A. Freund and L.M. Sehgal, Phys. Lett. {\bf B341} (1994) 90.
%
\bibitem{strat}
M. Stratmann and W. Vogelsang, Phys. Lett. {\bf B386} (1996) 370; \\
M. Stratmann and W. Vogelsang, Z Physik {\bf C74} (1997) 641.
%
\bibitem{hera}
A. De Roeck and T. Gehrmann, hep-ph/9711512 (1997).
%
\bibitem{twop}
S.J. Brodsky, hep-ph/9801288 (1998).
%
\bibitem{nlcg}
V. Telnov, hep-ph/9710014 (1997).
%
\bibitem{brod97}
S.J. Brodsky and I. Schmidt, Phys. Lett. {\bf B423} (1998) 145.
%
\bibitem{dhg}
S.D. Drell and A.C. Hearn, Phys. Rev. Lett. {\bf 162} (1966) 1520; \\
S.B. Gerasimov, Yad. Fiz. {\bf 2} (1965) 839.
%
\bibitem{moia}
S.D. Bass, Mod. Phys. Lett. {\bf A12} (1997) 1051.
%
\bibitem{low}
F. Low, Phys. Rev. {\bf 96} (1954) 1428; \\
M. Gell-Mann and M.L. Goldberger, Phys. Rev. {\bf 96} (1954) 1433.
%
\bibitem{brod69}
S.J. Brodsky and J.R. Primack, Ann. Phys. {\bf 52} (1969) 315.
%
\bibitem{brodsc}
S.J. Brodsky and I. Schmidt, Phys. Lett. {\bf B351} (1995) 344.
%
\bibitem{ansel}
M. Anselmino, B. L. Ioffe and E. Leader, Yad. Fiz. {\bf 49} (1989) 214.
%
\bibitem{gilman}
F.J. Gilman, Phys. Rev. {\bf 167} (1967) 1365.
%
\bibitem{hand}
L.N. Hand, Phys. Rev. {\bf 129} (1963) 1834.
%
\bibitem{bnt}
S.D. Bass, N.N. Nikolaev and A.W. Thomas,
Adelaide University preprint ADP-133-T80 (1990) unpublished; \\
S.D. Bass, Ph.D. thesis (University of Adelaide, 1992).
%
\bibitem{bint}
S.D. Bass, B.L. Ioffe, N.N. Nikolaev and A.W. Thomas,
J. Moscow Phys. Soc. {\bf 1} (1991) 317.
%
\bibitem{mank}
L. Mankiewicz and A. Sch\"afer, Phys. Lett. {\bf B242} (1990) 455; \\
L. Mankiewicz, Phys. Rev. {\bf D43} (1991) 64.
%
\bibitem{manoh}
A.V. Manohar, Phys. Rev. Lett. {\bf 66} (1991) 289.
%
\bibitem{lepage}
G.P. Lepage and S.J. Brodsky, Phys. Rev. {\bf D24} (1980) 2157.
%
\bibitem{ccm}
R.D. Carlitz, J.C. Collins, and A.H. Mueller, Phys. Lett. {\bf B214}
(1988) 229.
%
\bibitem{adler}
S.L. Adler, Phys. Rev. {\bf 177} (1969) 2426.
%
\bibitem{bell}
J.S. Bell and R. Jackiw, Nuovo Cimento {\bf 60A} (1969) 47.
%
\bibitem{maul}
M. Maul, B. Ehrnsperger, E.Stein and A. Sch\"afer,
Z Physik {\bf A356} (1997) 443.
%
\bibitem{mink} P. Minkowski, {\em in\/} Proc. Workshop on {\em
Effective Field Theories of the Standard Model\/}, Dobog\'{o}k\~{o},
Hungary 1991, ed. U.-G. Meissner (World Scientific, Singapore, 1992).
%
\bibitem{koeb}
R. K\"{o}berle and N.K. Nielsen, Phys. Rev. {\bf D8}
(1973) 660; \\
R.J. Crewther, S.-S. Shei and T.-M. Yan, Phys. Rev. {\bf D8} (1973) 3396;\\
R.J. Crewther and N.K. Nielsen, Nucl. Phys. {\bf B87} (1975) 52.
%
\bibitem{rjc} R.J. Crewther, Effects of Topological Charge in Gauge
Theories, in Facts and Prospects of Gauge Theories, Schladming,
Austria, February 1978, ed. P. Urban, Acta Physica Austriaca Suppl.
{\bf 19} (1978) 47, section 7.
%
\bibitem{collins} J. Collins, F. Wilczek and A. Zee, Phys. Rev.
{\bf D18} (1978) 242.
%
\bibitem{kod}
J. Kodaira, Nucl. Phys. {\bf B165} (1980) 129.
%
\bibitem{larin}
S.A. Larin, Phys. Lett. {\bf B334} (1994) 192.
%
\bibitem{jaffe}
R.L. Jaffe and A. Manohar, Nucl. Phys. {\bf B337} (1990) 509.
%
\bibitem{crewther}
R.J. Crewther, Riv. Nuovo Cimento {\bf 2} (1979) 63.
%
\bibitem{ioffe}
A.S. Gorskii, B.L. Ioffe, A.Yu. Khodjamirian,
Phys. Lett. {\bf B227} (1989) 474.
%


\end{thebibliography}
\end{document}